\documentclass[conference]{IEEEtran}
\usepackage[T1]{fontenc}
\usepackage[utf8]{inputenc}
\usepackage{amsmath,amssymb,bm}
\usepackage{graphicx}
\usepackage{booktabs}
\usepackage{tabularx}
\usepackage{geometry}
\usepackage{hyperref}
\hypersetup{colorlinks=true,allcolors=blue}
\usepackage{newunicodechar}
\newunicodechar{≈}{\approx}
\usepackage{mathtools}
\usepackage{amsthm}
\theoremstyle{definition}
\usepackage{algorithm}
\usepackage{algpseudocode}

\title{An Experimental Reservoir-Augmented Foundation Model: 6G O-RAN Case Study}

\author{
    \IEEEauthorblockN{
    Farhad Rezazadeh\IEEEauthorrefmark{1},
    Raymond Zhao\IEEEauthorrefmark{2},
    Jiongyu Dai\IEEEauthorrefmark{2},~Amir~Ashtari~Gargari\IEEEauthorrefmark{3}, Hatim~Chergui\IEEEauthorrefmark{4},\\ and Lingjia Liu\IEEEauthorrefmark{2}}
    \IEEEauthorblockA{\IEEEauthorrefmark{1}Hostelworld Group, Universitat Politècnica de Catalunya (UPC), Barcelona, Spain}
    \IEEEauthorblockA{\IEEEauthorrefmark{2}Virginia Tech, Blacksburg, VA, USA}
    \IEEEauthorblockA{\IEEEauthorrefmark{3}Centre Tecnol\'ogic de Telecomunicacions de Catalunya (CTTC/CERCA), Barcelona, Spain}
    \IEEEauthorblockA{\IEEEauthorrefmark{4}i2CAT Foundation, Barcelona, Spain}
    Email: farhad.rezazadeh@upc.edu,\{zraymond, jiongudai, ljliu\}@vt.edu, aashtari@cttc.es, hatim.chergui@i2cat.net}

\usepackage[font=small,labelfont=bf,labelsep=colon]{caption}
\begin{document}
\maketitle


\begin{abstract}
Next-generation open radio access networks (O-RAN) continuously stream tens of key performance indicators (KPIs) together with raw in-phase/quadrature (IQ) samples, yielding ultra-high-dimensional, non-stationary time series that overwhelm conventional transformer architectures. We introduce a reservoir-augmented masked autoencoding transformer (RA-MAT)\footnote{A minimal version of the source code (excluding ESN) is available at: \url{https://github.com/frezazadeh/Time-Series-Foundation-Model}}. This time series foundation model employs echo state network (ESN) computing with masked autoencoding to satisfy the stringent latency, energy efficiency, and scalability requirements of 6G O-RAN testing. A fixed, randomly initialized ESN rapidly projects each temporal patch into a rich dynamical embedding without backpropagation through time, converting the quadratic self-attention bottleneck into a lightweight linear operation. These embeddings drive a patch-wise masked autoencoder that reconstructs 30\% randomly masked patches, compelling the encoder to capture both local dynamics and long-range structure from unlabeled data. After self-supervised pre-training, RA-MAT is fine-tuned with a shallow task head while keeping the reservoir and most transformer layers frozen, enabling low-footprint adaptation to diverse downstream tasks such as O-RAN KPI forecasting. In a comprehensive O-RAN KPI case study, RA-MAT achieved sub-0.06 mean squared error (MSE) on several continuous and discrete KPIs. This work positions RA-MAT as a practical pathway toward real-time, foundation-level analytics in future 6G networks.
\end{abstract}


\section{Introduction}

The new 6G wireless technology will significantly improve communication by providing faster and more reliable connections \cite{frezazadehradio}, with O-RAN offering flexibility. However, setting up O-RAN comes with challenges \cite{Specialization_TVT} due to the large amounts of data it generates, such as time series data and performance indicators. Traditional machine learning (ML) methods struggle to handle this complex data effectively. Recent transformer and foundation models \cite{timeseries1, timeseries2} show promise in working with sequential data, but they face serious issues, such as high computational costs and a heavy need for large labeled datasets. These large, pre-trained systems perform well across different tasks with extra training. They learn mainly through unsupervised learning on big, unlabeled datasets, which allows them to build strong and versatile representations. To leverage these advantages, we introduce the RA-MAT, specifically designed to address the requirements of the 6G O-RAN environment. RA-MAT integrates ESN-based reservoir computing with masked autoencoding, thereby enhancing the efficiency and effectiveness of sequential data processing. Transformers require ESN integration due to their self-attention mechanisms, which, despite being powerful, lack the inherent capability to model fine-grained temporal dynamics efficiently. ESNs provide this capability by rapidly encoding temporal information into dynamic embeddings, thereby significantly reducing computational complexity and the need for extensive sequential data training. The primary contributions of this paper are as follows: 

\begin{table*}[]
\centering
\caption{Overview of statistical metrics for performance indicators observed in the experiments.}
\label{tab:kpi_stats}
\resizebox{1\linewidth}{!}{%
\begin{tabular}{c|c|c|c|c}
\hline
\textbf{KPI} &
  \textbf{Unit} &
  \textbf{Meaning} &
  \textbf{Observed Range} &
  \textbf{Mean $\pm$ Std Dev} \\ \hline
\textbf{Spectral Efficiency} &
  bps/Hz &
  Downlink data rate per unit of bandwidth \cite{oran-e2e} &
  0.00 – 3.74 &
  0.58 $\pm$ 0.38 \\
\textbf{RSRP} &
  dBm &
  Average reference-signal power per resource element at the UE &
  –102 – –75 &
  –87.58 $\pm$ 3.70 \\
\textbf{SINR} &
  dB &
  Ratio of received signal strength to interference plus noise &
  9.43 – 24.33 &
  18.31 $\pm$ 1.92 \\
\textbf{MIMO Rank} &
  — &
  Number of spatial data streams scheduled &
  1 – 2 &
  1.36 $\pm$ 0.38 \\
\textbf{MCS} &
  index &
  Chosen modulation and coding level for transport blocks &
  0 – 27 &
  9.04 $\pm$ 4.93 \\
\textbf{RB Number} &
  RBs &
  Count of physical resource blocks allocated &
  2 – 25 &
  22.31 $\pm$ 4.34 \\
\textbf{CQI} &
  index &
  UE’s wideband channel quality feedback &
  0 – 13 &
  8.51 $\pm$ 0.92 \\
\textbf{RSRQ} &
  dB &
  Composite metric of signal and total received power quality &
  –14.00 – –6.40 &
  –10.55 $\pm$ 2.47 \\
\textbf{PMI} &
  index &
  Preferred precoding index reported by the UE &
  0 – 3 &
  0.93 $\pm$ 0.84 \\
\textbf{UE RSSI} &
  dBm &
  Total received power level at the UE &
  –70 – –60 &
  –65.36 $\pm$ 2.62 \\
\textbf{UE Buffer Status} &
  bytes &
  Volume of data queued in the UE uplink buffer &
  0 – 2944 &
  25.61 $\pm$ 85.34 \\
  \textbf{BLER} &
  \% &
  Rate of block transmission errors &
  0 – 78.00 &
  2.64 $\pm$ 6.76 \\
\textbf{Packet Delay} &
  ms &
  Round-trip time from server to client per packet &
  0 – 3048.06 &
  62.70 $\pm$ 208.87 \\
 \hline
\end{tabular}%
}
\end{table*}

\begin{itemize} 
\item We introduce the RA-MAT architecture, which combines reservoir computing and masked autoencoding transformers to enable efficient processing of ultra-high-dimensional, non-stationary time series data typical in 6G O-RAN systems. 
\item By front‐loading each temporal patch with a fixed, randomly initialized reservoir, we endow every patch embedding with rich, short‐term dynamic memory, without incurring the computational or memory overhead of backpropagation through time. This lightweight recurrent encoder relieves the transformer’s self-attention layers from having to re-learn fine-grained temporal correlations, enabling us to drastically reduce model size and training cost while preserving fidelity on ultra-high-dimensional, non-stationary 6G O-RAN KPI streams. In effect, the ESN serves as a zero-cost temporal inductive bias that transforms quadratic attention bottlenecks into efficient linear projections.
\item Based on thorough testing with real-world O-RAN KPI data, we show that RA-MAT excels in tasks like KPI forecasting.
\end{itemize}

\section{Data Collection and Preprocessing}
The data are collected from the O-RAN-based video streaming experiments over 10 days.
For each experimental round, the streaming request on the UE side lasts for 120 seconds. The testbed automatically collects physical and MAC-layer statistics from the base station and UE, Wireshark packet captures, FFmpeg statistics logs, and video recordings as raw data to construct our KPI dataset. Physical and MAC-layer statistics are recorded every 20 ms with corresponding timestamps and saved into a .csv file\footnote{A minimal version of our dataset is available at: \url{https://ieee-dataport.org/documents/video-streaming-network-kpis-o-ran-testing}}. An overview of the KPIs observed in the experiments, including their units, meanings, observed ranges, and statistical metrics, is summarized in Table \ref{tab:kpi_stats} \cite{dataset_oran}.

\subsubsection{Calculate the moving average}
Due to the collection of raw data from various components with differing timestamps, preprocessing is essential to synchronize them. We employ a fixed-length time window to compute the moving average of the raw data.
Specifically, for each KPI, if multiple raw measurements share timestamps within a given time window, we calculate their average and assign the window’s start time, $t_{start}$, as the timestamp for this average value. This approach facilitates data alignment.
After processing all KPIs within the current time window, we advance the window by a fixed step length, updating the start time to $t_{start}+t_{step}$. This averaging process is repeated for the entire dataset.
Once the time window’s end time reaches the experiment’s conclusion, we obtain a sequence of averaged values with corresponding timestamps for each KPI. These sequences are then aggregated into rows in our dataset, $D_{kpi}$, for further analysis. Notably, some rows may contain missing values for certain KPIs.

\subsubsection{Padding and Filtering}
To prepare the datasets $D_x$ and $D_y$ for training and testing, we first inspect $D_{kpi}$ for rows with missing values. If a row lacks only the ‘UE Packet Delay’ value, we impute this missing value with -1 to distinguish it from typical values. Conversely, if any other KPI values are missing, the entire row is removed. To ensure consistency and reasonable data, Inter-quartile Range (IQR) pruning was used to remove outliers from the dataset. We chose to set the range to the $Q1$=10th and $Q3$=90th quartiles and $IQR = Q3-Q1$. The corresponding thresholds to keep were determined by $lower=Q1-1.5\times IQR$ and $upper=Q3+1.5\times IQR$. We determined whether pruning was necessary by plotting the time series data to identify any data points significantly beyond standard ranges, and then comparing them to the average and variance of the features.  

\subsubsection{Construct sequential data}
Assuming a required sequence length of $N_{seq}$, we iterate through the rows of $D_{kpi}$ post-dropping and padding, starting from the $N_{seq}$-th row. For the $i$-th row, we first verify if the timestamp difference between the $i$-th and $(i+1)$-th rows equals $t_{step}$. If so, we examine the sequence from the $i$-th to the $(i-N_{seq}+1)$-th row to ensure all consecutive timestamp differences are $t_{step}$. If this condition is met, we stack the KPI vectors from the $i$-th to the $(i-N_{seq}+1)$-th row to form a data point in dataset $D_x$, with the KPI vector from the $(i+1)$-th row serving as the corresponding data point in dataset $D_y$.
Through this process, we construct $D_x$ with dimensions $X=[M, N_{seq}, K]$ and $D_y$ with dimensions $y=[M, K]$, where $M$ denotes the total number of data points and $K$ represents the number of KPI features.

\begin{figure*}[t]
\centering
\includegraphics[width=2\columnwidth, clip,trim={0cm 0cm 0cm 0cm}]{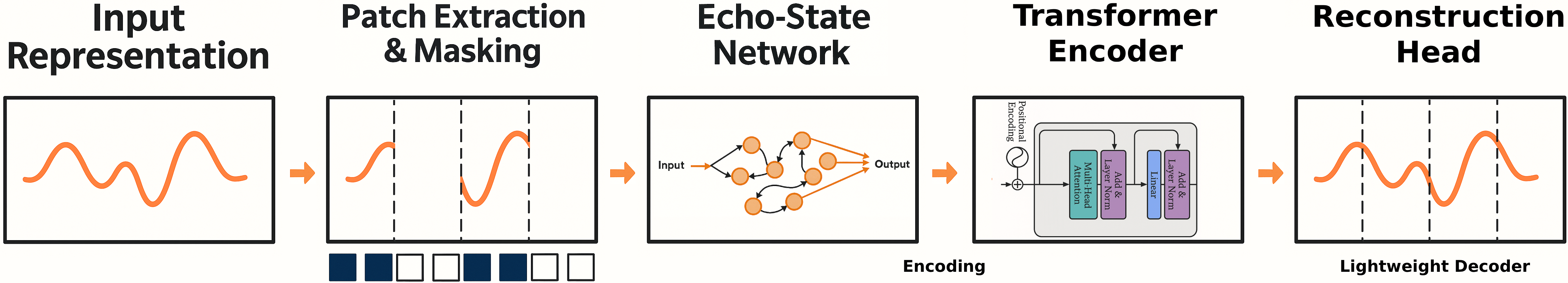}
 \caption{Overview of the RA-MAT Five-Stage Pipeline. From left to right, a single multivariate time series is (1) represented as a raw input window, (2) split into non-overlapping temporal patches with a subset randomly masked, (3) passed through a fixed ESN to produce dynamic embeddings, (4) encoded via a stack of transformer layers (with added positional embeddings) to capture both local and global structure, and (5) decoded by a lightweight reconstruction head that predicts only the masked patches.}
\label{fig:framework}
\end{figure*}

\section{Two-Stage Self-Supervised and Supervised Fine-Tuning Framework}

The methodology outlined describes a detailed two-stage process. The first stage is an unsupervised pre-training phase, where the focus is on creating general-purpose representations. This is followed by a supervised fine-tuning phase designed to specialize the model in the prediction or classification of specific KPIs. The pseudocode for the unsupervised pre-training and the supervised fine-tuning workflows is summarized in Algorithm \ref{alg:mae-pretrain-descr} and Algorithm \ref{alg:finetune-descr}, respectively.
\begin{algorithm}[t!]
\caption{Pretraining of RA-MAT}
\label{alg:mae-pretrain-descr}
\small
\begin{algorithmic}[1]
\Require
  \textit{raw\_series} \Comment{Collection of multivariate time series}\\
  \textit{window\_length}, \textit{patch\_length}, \textit{mask\_ratio}\\
  \textit{embed\_dim}, \textit{reservoir\_size}, \textit{num\_layers}\\
  \textit{initial\_learning\_rate},\textit{weight\_decay}, \textit{total\_steps}, \textit{warmup\_steps}
\State Initialize all model parameters, the fixed ESN reservoir, the optimizer, and the LR scheduler
\For{each training epoch}
  \For{each time series in \textit{raw\_series}}
    \State \textbf{Standardize KPIs:}
      compute KPI means and variances over the full series;
      subtract the mean and divide by the standard deviation for each value
    \For{each sliding window of length \textit{window\_length}}
      \State \textbf{Divide window into patches:}
        split into non-overlapping segments of length \textit{patch\_length}; flatten each to a vector
      \State \textbf{Select masked patches:}
        for each patch, flip a biased coin with probability \textit{mask\_ratio}; collect mask positions
      \State \textbf{Run ESN reservoir:}
        feed patches in sequence into the ESN to get hidden states
      \State \textbf{Project to embedding space:}
        apply a linear layer to each reservoir state and add a positional offset
      \State \textbf{Insert mask tokens:}
        replace embeddings at masked positions with a shared trainable mask token
      \State \textbf{Encode with transformer:}
        pass the full sequence of embeddings through \textit{num\_layers} of self-attention + feed-forward blocks
      \State \textbf{Decode masked patches:}
        for each masked position, map the corresponding transformer output back to patch space
      \State \textbf{Compute loss:}
        mean squared error between reconstructed and original masked patches
      \State \textbf{Update model:}
        backpropagate loss, clip gradients, take an optimizer step, update learning rate
    \EndFor
  \EndFor
\EndFor
\end{algorithmic}
\end{algorithm}

\begin{algorithm}[t!]
\caption{Supervised Fine-Tuning of RA-MAT}
\label{alg:finetune-descr}
\small
\begin{algorithmic}[1]
\Require
  \textit{pretrained\_encoder}, labeled pairs \textit{(input, target)}\\
  \textit{batch\_size}, \textit{num\_epochs}, per-layer learning rates, freeze mode
\Ensure
  fine-tuned encoder and prediction head
\State Initialize prediction head (either linear for regression or softmax for classification)
\State Select which parts of the encoder are trainable based on \textit{freeze mode}
\For{each epoch from 1 to \textit{num\_epochs}}
  \State Shuffle the labeled data
  \For{each mini-batch of size \textit{batch\_size}}
    \State \textbf{Forward pass:}
      feed batch through encoder;
      aggregate token embeddings by averaging across time
    \If{regression task}
      \State compute predictions with the linear head
      \State compute mean squared error loss
    \Else
      \State compute class probabilities with the softmax head
      \State compute cross-entropy loss
    \EndIf
    \State \textbf{Backward pass:}
      compute gradients, clip if necessary
    \If{using layer-wise learning rates}
      \State update each layer with its assigned rate
    \Else
      \State update all trainable parameters with the same rate
    \EndIf
  \EndFor
  \State Optionally validate on held-out data
\EndFor
\end{algorithmic}
\end{algorithm}

During the pre-training phase, large sets of raw, unlabeled time series data\footnote{\url{https://huggingface.co/datasets/AutonLab/Timeseries-PILE}} in CSV format undergo initial standardization and KPI padding. The data is then segmented into fixed-length windows and divided into non-overlapping temporal patches. A strategic masking technique is applied, masking about 30\% of these patches randomly, while the unmasked patches are encoded using an ESN. This technique produces high-dimensional dynamic embeddings, eliminating the need for backpropagation through time. The reservoir outputs are then projected into a lower-dimensional token space and augmented with positional embeddings. Masked patches are replaced with a trainable mask token, and several transformer encoder layers process the resulting sequence. These layers use multi-head self-attention and feed-forward networks to learn both local and global patterns in the time series. A simple decoder reconstructs the masked patches, and the reconstruction error is measured by MSE. The model is trained using the AdamW optimizer, with gradient clipping and a cosine learning rate schedule.
\begin{figure*}[t]
\centering
\includegraphics[width=2\columnwidth, clip,trim={0cm 0cm 0cm 0cm}]{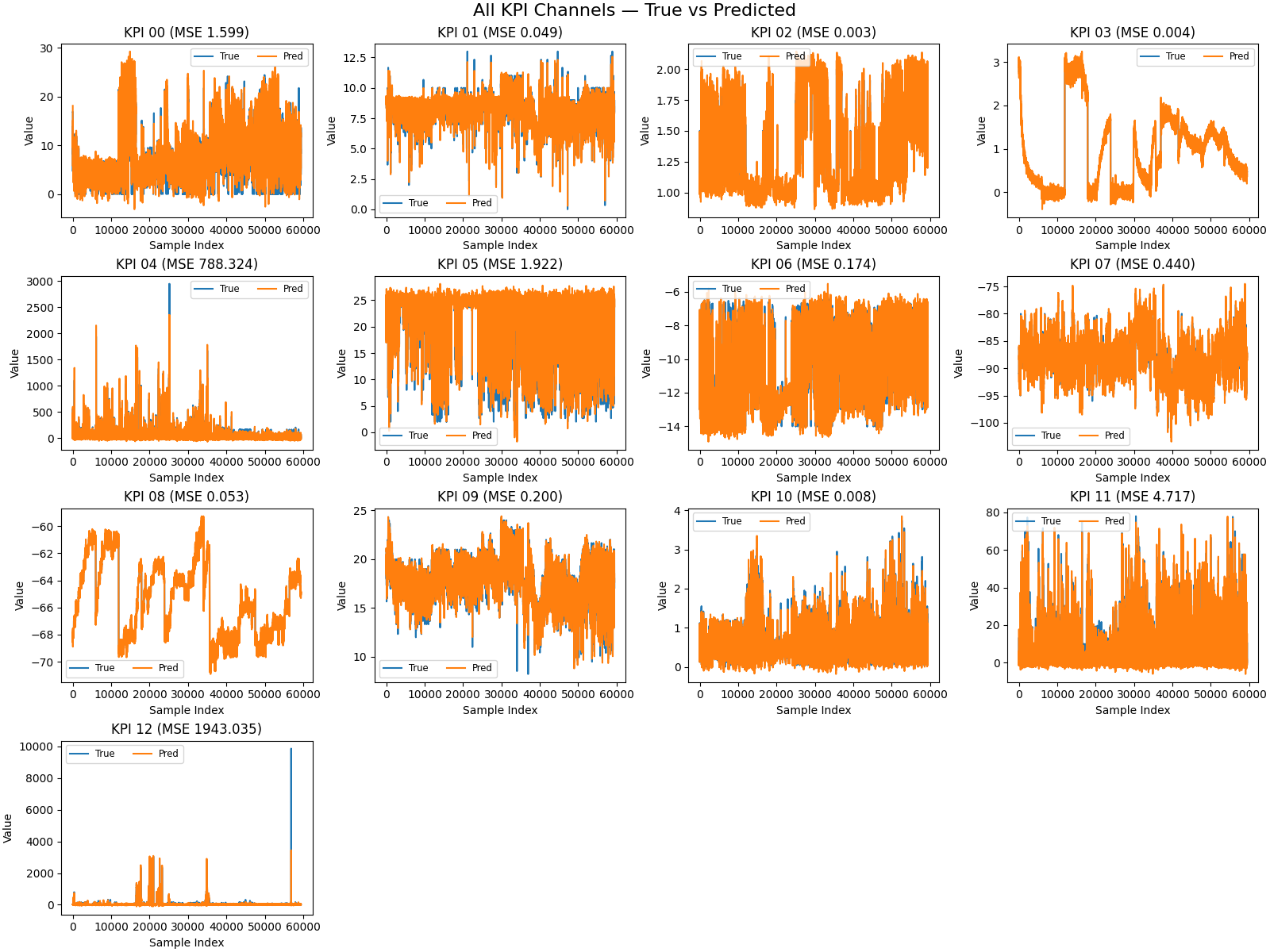}
 \caption{Initial inference results on O-RAN KPIs.}
\label{fig:performance}
\end{figure*}

In the fine-tuning phase, the pre-trained encoder weights are initialized, and per-channel standardization is applied to both inputs and targets using pre-fitted scalers. A lightweight task-specific head is attached to the encoder for either regression or classification tasks. During this phase, the reservoir and initial transformer layers remain static, with training focused on the top transformer blocks and the appended head. The model optimization follows either MSE for regression tasks or cross-entropy loss for classification tasks, still using AdamW, along with options for layer-wise learning rate decay and gradient clipping. The fine-tuning process includes early stopping based on validation performance. The result is a final, fine-tuned model, along with the necessary scalers for real-time deployment, ensuring low-latency and high-accuracy inference suitable for applications in 6G O-RAN.

The Figure \ref{fig:framework} illustrates our framework workflow: the first four blocks (Encoding) transform the masked-patch sequence into rich contextual embeddings. In contrast, the final block (Lightweight Decoder) reconstructs only the held-out patches. By freezing the ESN and most transformer layers during fine-tuning, RA-MAT delivers an efficient foundation model for real-time KPI forecasting in 6G O-RAN.

\section{Inference Performance Across All KPIs}
Figure~\ref{fig:performance} shows the model’s one‐step–ahead predictions (orange) against the ground truth (blue) for all 13 KPIs, with the corresponding test‐set MSE printed in each subplot header. Several trends are immediately apparent:

\begin{itemize}
  \item \textbf{High-fidelity reconstruction (MSE<0.06):}  
    KPIs 2 (SINR, MSE=0.003), 3 (MIMO rank, MSE=0.004), 8 (PMI, MSE=0.053), and 10 (UE Buffer Status, MSE=0.008) exhibit almost perfect overlap of true and predicted curves. These metrics have relatively stable ranges and either low complex noise (SINR/RSRQ-type measures) or small discrete value sets (rank, PMI), making them easier for the transformer to model.

  \item \textbf{Moderate-error KPIs (MSE$≈$0.05-2):}  
    KPIs 0 (spectral efficiency, MSE=1.599, 1 (RSRP, MSE=0.049), 5 (RB count, MSE=1.922), 7 (UE RSRQ, MSE=0.440), and 9 (UE RSSI, MSE=0.200) show somewhat larger deviation during rapid fluctuations or burst events. In these plots, the predictions track the overall trend but lag the high-frequency spikes in the true signal, suggesting that local irregularities, such as sudden scheduling changes or buffer overflows, are more challenging to capture.

  \item \textbf{High-variance, heavy-tail KPIs (MSE$>$2):}  
    KPIs 4 (MCS index, MSE=788.324), 11 (BLER, MSE=4.717), and 12 (packet delay, MSE=1943.035) suffer the most significant errors. These metrics either span broad dynamic ranges (packet delay up to multiple seconds, BLER spikes up to 78 \%) or exhibit very sparse, bursty behaviour (MCS changes in rapid retransmission scenarios). As a result, the model’s mean‐squared loss is dominated by occasional extremes that are underrepresented in the training set.

\item \textbf{Impact of outliers:}  
    Our retention of extreme but valid KPI values (e.g.\ rare packet‐delay spikes or transient MCS jumps) in the test set inflates the MSE on certain KPIs. We deliberately kept these outliers to reflect real‐world O-RAN behavior in our evaluation transparently.

 \item \textbf{Unified continuous-discrete support:}  
    Importantly, RA-MAT’s reservoir-augmented embedding and flexible output head allow seamless handling of both continuous‐valued KPIs (e.g.\ RSRP, packet delay) and discretized indices (e.g.\ MIMO rank, MCS, PMI) without architectural changes.
  
  \item \textbf{Qualitative trends:}  
    Despite the elevated MSE in several KPIs, the predicted curves generally follow the correct long‐term trends. In particular, during steady-state periods, the model almost perfectly aligns with the true values, and its most considerable divergences coincide with transient events where even human‐tuned heuristics would struggle.
\end{itemize}

Overall, Figure~\ref{fig:performance} demonstrates that RA-MAT captures both slow‐varying baselines (e.g.\ signal strength, resource allocation) and recurring patterns (e.g.\ rank/MCS scheduling), while infrequent, high‐amplitude KPI excursions primarily drive its reconstruction error. These insights will guide future work on loss reweighting or data‐augmentation strategies to better handle heavy‐tailed KPI distributions.

\section{Conclusion}
In this work, we introduced RA-MAT, a novel reservoir-augmented masked autoencoding transformer tailored for ultra-high-dimensional, non-stationary time series data streams in 6G O-RAN environments. By integrating a fixed ESN with a masked autoencoding pre-training objective, RA-MAT overcomes the quadratic self-attention bottleneck and obviates backpropagation through time, yielding a lightweight yet expressive foundation model. Through comprehensive experiments on real-world O-RAN KPI datasets, we demonstrated that RA-MAT achieves sub-0.06 MSE on several stable KPIs while maintaining competitive performance on bursty, heavy-tailed metrics. Our two-stage training framework, self-supervised pre-training followed by low-footprint fine-tuning, enables rapid adaptation to diverse downstream tasks such as KPI forecasting, anomaly detection, and fault classification, all under the stringent latency and energy constraints of next-generation networks. In the future, we will enhance RA-MAT by developing adaptive masking and loss-reweighting strategies to capture rare, high-impact KPI events better; integrating multi-modal fusion of raw IQ samples and packet-level features for richer, cross-domain embeddings; designing hierarchical, multi-scale ESNs to model fast transients and long-term trends simultaneously, and broadening evaluation to new 6G use cases, such as mobility prediction, dynamic beamforming, and end-to-end quality-of-experience modeling, to establish RA-MAT as a versatile, scalable foundation for next-generation wireless analytics and control.

\bibliographystyle{IEEEtran}

\end{document}